\documentclass[aps,pra,twocolumn,floatfix,showpacs,lengthcheck,citeautoscript,superscriptaddress]{revtex4-1}
\usepackage{graphicx}
\usepackage{amsmath}
\usepackage{amsfonts}
\usepackage{mathrsfs}
\usepackage{color}

\DeclareMathOperator{\sgn}{sgn}

\usepackage{amssymb}
\usepackage{bm}
\usepackage{times}
\usepackage{MnSymbol}

\begin{document}

\title{Population inversion and dynamical phase transitions in a driven superconductor}

\author{H. P. {Ojeda Collado}} 
\affiliation{Centro At\'omico Bariloche and Instituto Balseiro,
  Comisi\'on Nacional de Energ\'ia At\'omica, 8400 Bariloche,
  Argentina}
\affiliation{Consejo Nacional de Investigaciones Cient\'ificas y
  T\'ecnicas (CONICET), Argentina}
\author{Jos\'{e} Lorenzana}
\email[Corresponding author: ]{jose.lorenzana@cnr.it}
\affiliation{ISC-CNR and Department of Physics, Sapienza University of Rome, Piazzale Aldo Moro 2, I-00185, Rome, Italy}
\author{Gonzalo Usaj}
\affiliation{Centro At\'omico Bariloche and Instituto Balseiro,
  Comisi\'on Nacional de Energ\'ia At\'omica, 8400 Bariloche,
  Argentina} 
\affiliation{Consejo Nacional de Investigaciones Cient\'ificas y
  T\'ecnicas (CONICET), Argentina}
\author{C. A. Balseiro}
\email[Corresponding author: ]{balseiro@cab.cnea.gov.ar}
\affiliation{Centro At\'omico Bariloche and Instituto Balseiro,
  Comisi\'on Nacional de Energ\'ia At\'omica, 8400 Bariloche,
  Argentina}
\affiliation{Consejo Nacional de Investigaciones Cient\'ificas y
  T\'ecnicas (CONICET), Argentina}

\begin{abstract}
We consider a superconductor in which the density of states at the Fermi level or the pairing interaction is driven periodically with a frequency larger than the 
superconducting gap in the collisionless regime.  We show by numerical and analytical computations that a subset of quasiparticle excitations enter into resonance and perform synchronous Rabi 
oscillations leading to cyclic population inversion with a frequency that depends on the amplitude of the drive. As a consequence a new ``Rabi-Higgs'' mode emerges. Turning off the drive at different times and modulating the 
strength allows access to all known dynamical phases of the order parameter: persistent oscillations, oscillations with damping and overdamped dynamics. We discuss physical realizations of the drive and methods to detect the dynamics. 
\end{abstract}
\date{\today}
\maketitle

Quasiparticle relaxation times in superconductors can ea\-si\-ly reach nanoseconds at low temperatures \cite{dynes1978} while the typical order parameter dynamics is on the hundreds of femtoseconds time scale ~\cite{Mansart2013,Lorenzana2013,Matsunaga2013,Matsunaga2014,Cea2016b}, leaving a large window where, in principle, it is possible to observe energy-conserving, out-of-equilibrium dynamics of the superconducting order parameter.
Ultra-cold Fermi gases provides another platform where the order parameter  dynamics of a fermionic condensate can be studied in real time thanks to the possibility to couple directly to it~\cite{Behrle2018}, or indirectly through the modulation of Feshbach resonances \cite{Chin2010,Clark2015}, 
and external potentials defined by optical traps \cite{Bloch2008}. 

Restricting to Bardeen, Cooper and Schrieffer (BCS) like wave-functions and within BCS reduced Hamiltonian, it has been shown \cite{Barankov2004,Barankov2006a,Yuzbashyan2006} that there are essentially three zero-temperature dynamical phases, 
characterized by the qualitatively different time evolution of the order parameter after a sudden change of the interaction. For small perturbations of the BCS ground state, one finds an oscillatory behavior with a frequency
$2\Delta_\infty$ and a $t^{-\frac{1}{2}}$ decay of the order parameter to a steady state value $\Delta_{\infty}$, smaller than the thermodynamical order parameter $\Delta_0$ at equilibrium. This can be understood appealing to a linearized dynamics 
where quasiparticle excitations evolve with a frequency determined by its own quasiparticle energy leading to the dephasing of the excitations that build the perturbation and washing out any macroscopic manifestation of the dynamics. On the other hand, 
if the perturbation is large, two outcomes are possible. If one starts from an initial condition where the order parameter is much larger than the
equilibrium value, the dynamics becomes overdamped 
and the asymptotic stationary  value becomes zero.  If instead, one starts from an order parameter much smaller than the one at equilibrium,  persistent oscillations occur where all 
quasiparticles evolve synchronously, driven by the pairing field that they build self-consistently \cite{Barankov2006a}. 

In the case of an isotropic band structure, these spontaneous amplitude modes of the superconducting gap (Higgs-like modes) are totally symmetric, {\em i.e.} there is no dependence of the modes on the Euler angles parameterizing a point on the 
Fermi surface. More importantly, because of the approximate particle-hole symmetry of the BCS state, the charge is locally conserved in each real space unit cell and therefore long-range effects of the Coulomb interactions are irrelevant and can be neglected from the outset.  

In this work we study the effect of a periodic drive\cite{Tsuji2015,Sentef2017} that couples to the amplitude of the superconducting order parameter. We discuss several type of drives and how they can be rea\-li\-zed in practice. In principle, different drives can be implemented 
in ultra-cold atoms where the easy of manipulation is one of their well known characteristics \cite{Chin2010,Bloch2008}.   
In solid state systems, on the other hand,  one would think that driving requires more effort. However, the first observations of the out of equilibrium fermionic condensate dynamics were done in such systems \cite{Mansart2013,Matsunaga2013,Matsunaga2014}. Drives that modulate the density of states (DOS) assisted by phonons have been proposed long ago \cite{Balseiro1980,Littlewood1982}. 
We argue that in unconventional superconductors also phonon assisted drives that modulate the coupling constant are feasible. Other proposed realizations are impulsed stimulated Raman scattering (ISRS) drives \cite{Mansart2013,Lorenzana2013} and THz drives \cite{Xi2013,Matsunaga2013,Matsunaga2014,Cea2016b,Papenkort2007,Papenkort2008,Krull2014}. 
We find that under such periodic drives (both for ultra-cold atomic gases and solid state systems) a subset of quasiparticles, whose energy matches the frequency of the drive, enter into resonance and perform synchronous Rabi oscillations resulting in an oscillation of the 
amplitude which will be referred to as ``Rabi-Higgs'' mode that is clearly distinct from the spontaneous Higgs mode discussed before in the literature \cite{Barankov2004,Barankov2006a,Yuzbashyan2006}. 

An important physical consequence of the Rabi-Higgs mode is that a periodic population inversion is produced for the resonant quasiparticles, with the period determined by the amplitude of the drive. This result  can be understood mapping the wave-function to a system of Anderson pseudo-spins and applying the rotating wave approximation (RWA) in analogy with Rabi oscillations in nuclear magnetic resonance (NMR) experiments. 
We show that it is possible to explore the full dynamical phase diagram of an out-of-equilibrium fermionic condensate by turning off the drive at different instants of the Rabi cycle, very much as in regular pulsed NMR experiments.

In addition to the neutral spontaneous amplitude modes one can consider also charge modes. The simplest spontaneous charge mode is the longitudinal sound mode which in superconductors is pushed to the plasma frequency by the Anderson-Higgs 
mechanism \cite{Anderson1958}. There are also zero-momentum spontaneous charge modes which leave the unit cell neutral but consist of charge fluctuations within the cell. Typically these charge modes are Raman active and produce decaying oscillatory 
responses with frequency $2\Delta_0/\hbar$ for small perturbations (rather than the plasma frequency) because they do not involve the long-range Coulomb interaction \cite{Devereaux1995,Cea2016b,Maiti2017}. These are the 
oscillations observed experimentally in real time in Ref.~\cite{Mansart2013}. In the following, for the sake of simplicity, we will restrict to drives that couple to the 
pure amplitude modes although it will become clear that similar physics will
emerge for more general Raman drives that can couple with intra-cell charge fluctuations.


\section{Model \label{sec:model}}
Our goal is analyze the non-equilibrium superconducting response under the presence of an uniform time-dependent drive of the condensate. 
For simplicity we consider a BCS single-band s-wave superconductor. 
We describe the dy\-na\-mics of the superconductor in terms of Anderson pseudo-spins by using a time-dependent BCS hamiltonian,
\begin{equation}\label{eq:ham}
H=-2\sum_{\bm{k}} \xi_{\bm{k}}(t)S_{\bm{k}}^{z}-\lambda(t)\sum_{\bm{k},\bm{k}'}S_{\bm{k}}^{+}S_{\bm{k}'}^{-}\,,
\end{equation} 
where $\xi_{\bm{k}}(t) =\varepsilon_{\bm{k}}(t)-\mu$, $\varepsilon_{\bm{k}}(t)$ is the free particle energy, $\mu$ is the Fermi level and $\lambda(t)$ is the pairing interaction. We allow both parameters to be time-dependent although we will  analyze their effects separately.  
The possible physical realizations of these drives will be discussed in the Section II. 
The pseudo-spin operators are given by,
\begin{eqnarray}
\nonumber
S_{\bm{k}}^{x}&=&\frac{1}{2}\left(c_{\bm{k}\uparrow}^{\dagger}c_{-\bm{k}\downarrow}^{\dagger}+c_{-\bm{k}\downarrow}^{}c_{\bm{k}\uparrow}^{}\right)\,, \\ 
\nonumber
S_{\bm{k}}^{y}&=&\frac{1}{2i}\left(c_{\bm{k}\uparrow}^{\dagger}c_{-\bm{k}\downarrow}^{\dagger}-c_{-\bm{k}\downarrow}^{}c_{\bm{k}\uparrow}^{}\right)\,, \\
\label{eq:sz}
S_{\bm{k}}^{z}&=&\frac{1}{2}\left(1-c_{\bm{k}\uparrow}^{\dagger}c_{\bm{k}\uparrow}^{}-c_{-\bm{k}\downarrow}^{\dagger}c_{-\bm{k}\downarrow}^{}\right)\,,
\end{eqnarray}
and $S_{\bm{k}}^{\pm}\equiv S_{\bm{k}}^{x}\pm iS_{\bm{k}}^{y}$ is the usual ladder operator. Here $c^\dagger_{\bm{ k}\sigma}$ ($c_{\bm{k}\sigma}$) is the creation (destruction) operator for electrons with momentum $\bm{k}$ and spin $\sigma$. 

Due to the infinite range of interactions, assumed in the second term of Eq.~(\ref{eq:ham}), the mean-field approximation is exact in the thermodynamic limit and the time-dependent mean-field equations describe the exact dynamics. 
The BCS mean-field Hamiltonian can be written as
\begin{equation}\label{eq:hamMF}
H_{\mathrm{MF}}=-\sum_{\bm{k}}\bm{S}_{\bm{k}}\cdot\bm{b}_{\bm{k}},
\end{equation}
where $\bm{b}_{\bm{k}}\left(t\right)=2\left(\Delta'\left(t\right),\Delta''\left(t\right),\xi_{\bm{k}}(t) \right)$ is an effective magnetic field. The real and imaginary part of the instantaneous superconducting order parameter are defined self-consistently as
\begin{eqnarray}
\Delta'\left(t\right)&=&\lambda(t)S_{t}^x,
\label{eq:del1}\\ 
\Delta''\left(t\right)&=&\lambda(t)S_{t}^y, \label{eq:del2}
\end{eqnarray}
with $S_t^x\equiv\sum_{\bm{k}}\left\langle S_{\bm{k}}^{x}\right\rangle$ and $S_t^y\equiv\sum_{\bm{k}}\left\langle S_{\bm{k}}^{y}\right\rangle$.
Clearly the $x$-$y$ projection of the pseudo-spins are related to the superconducting order parameter while the $z$ projection is related to charge fluctuations through Eq.~(\ref{eq:sz}).

From hereon we will denote equilibrium static quantities with a ``$0$'' superscript$/$subscript. Without loss of generality we take $\left\langle S_{\bm{k}}^{y}\right\rangle^{0}=0$. 
At equilibrium, in the absence of excitations, the pseudo-spins align in the direction of their local field $\bm{b}_{\bm{k}}^0=2\Delta_{0}\,\hat{\bm{x}}+2\xi_{\bm{k}}\,\hat{\bm{z}}$ in order to minimize the system's energy, which is
described by the spin Hamiltonian [Eq.~(\ref{eq:hamMF})]. The zero-temperature pseudo-spin texture is given by
\begin{equation}
\label{eq:sequilib} 
\left\langle S_{\bm{k}}^{x}\right\rangle^{0}=\frac{\Delta_{0}}{2\sqrt{\xi_{\bm{k}}^{2}+\Delta_{0}^{2}}},\qquad \left\langle S_{\bm{k}}^{z}\right\rangle^{0}=\frac{\xi_{\bm{k}}}{2\sqrt{\xi_{\bm{k}}^{2}+\Delta_{0}^{2}}}.
\end{equation}
Out of equilibrium the Anderson pseudo-spins obey the equations of motion for magnetic moments in a time-dependent magnetic field,
\begin{equation}
\label{eq:eom}
\frac{d\langle\bm{S}_{\bm{k}}\rangle}{dt}=-\bm{b}_{\bm{k}}\left(t\right)\times\langle\bm{S}_{\bm{k}}\rangle,
\end{equation}
where $\hbar\equiv1$ as we assume for the rest of the paper.

It is simple to show that momentum independent modulations of the quasi-particle energy are irrelevant. Indeed, consider such a time-dependent modulation in impulsive form, 
\begin{equation}
\label{eq:drivecte}
\varepsilon_{\bm{k}}(t)= \varepsilon_{\bm{k}}^0+ V\Delta t \thinspace \delta(t), 
\end{equation}
where  $\varepsilon_{\bm{k}}^0$ is the equilibrium dispersion relation and $V\Delta t$ is the strength of an impulsive potential that couples to the total charge. Integrating the equation of motion in a small interval of 
time $dt$ we find that after the impulse the pseudo-spins obey
\begin{equation}
\langle\bm{S}_{\bm{k}}\rangle(dt)=\langle\bm{S}_{\bm{k}}\rangle ^0-  
\hat{\bm{z}} \times\langle\bm{S}_{\bm{k}}\rangle ^0 V\Delta t
\end{equation}
where $\langle\bm{S}_{\bm{k}}\rangle ^0$ is the equilibrium pseudo-spin before the impulse. This corresponds to a global rotation of all pseudo-spins around the
$z$-axis by the same angle $\Delta \phi= - V \Delta t$ which translates through Eqs.~(\ref{eq:del1}) and (\ref{eq:del2}) in a global rotation of the order parameter by the same angle in the $x$-$y$ plane. Therefore, after the
perturbation the pseudo-spins are still in equilibrium and the only consequence is a change of the global phase of the superconductor. For an isolated superconductor such change has no physical consequences and can be gauged away.
This is in agreement with the analysis of Raman scattering \cite{Cea2016} which shows that a Raman operator proportional to the total density does not produce 
scattering regardless of whether one considers long-range interactions or not. 
As a consequence, uniform (momentum independent) drives as the one in Eq.~(\ref{eq:drivecte}) can be eliminated from the outset.
For the same reason we can assume that $\mu$ is time independent and corresponds to the equilibrium chemical potential before the drive. Also, for the same reason, 
long-range Coulomb interactions do not play any relevant role for this class of drives. Long-range Coulomb interactions may become relevant if one considers drives that couple to the density operator at finite momentum which is beyond the scope of this work. 



\section{Driving Mechanisms of the Superconducting Condensate}
In this section we discuss different mechanisms that can be used to drive the superconductor out of equilibrium. Contrary to previous works in the context of interaction quenches in ultracold-atomic systems \cite{Barankov2004,Barankov2006a,Yuzbashyan2006,Yuzbashyan2006a,Bunemann2017}, we shall consider a periodic time-dependent perturbation acting over a long time. 

\subsection{Phonon-assisted density of states driving}
The study of the coupling of phonons to the spontaneous amplitude mode of the order parameter started several decades ago in relation with $2$H-NbSe$_2$ \cite{Balseiro1980,Littlewood1982,Browne1983,Cea2014}. In this material a charge-density-wave appears at an ordering temperature $T_{\mathrm{CDW}}=33$ $^{\circ}$K above the superconducting critical temperature, $T_\mathrm{c}=7$ $^{\circ}$K. The CDW partially gaps the Fermi surface in such a way that a Raman phonon, which drives the CDW,  
strongly modulates the density of states (DOS) at the Fermi level available for superconductivity. As a consequence, at low temperatures in the superconducting phase the equilibrium amplitude of the order parameter is strongly 
dependent on the lattice coordinate, which we will denote as $u$. 

Since the relevant phonon is Raman active it can be launched impulsively \cite{Merlin1997,Stevens2002}. Assuming that the light pulse is applied at $t=0$ and that damping is negligible, the phonon coordinate obeys 
$u(t)=u_0\,\vartheta\left(t\right)\sin\left(\omega_d t\right)$ where $\vartheta\left(t\right)$ is the Heaviside step function and $\omega_d$ denotes the driving frequency (in this case the
phonon frequency).  
A change in the DOS can be introduced as a change in the Fermi velocity which corresponds to a time dependence in Eq.~(\ref{eq:ham}) of the form 
$\varepsilon_{\bm{k}}(t)=\varepsilon_{\bm{k}}^0[1+\beta(t)]$ with $\beta(t)\propto u(t)$.
Because, as discussed above,  dynamical terms that couple to the total density are irrelevant, we can add a time-dependent term proportional to the chemical potential so that the relevant perturbation is given by 
\begin{equation}
  \label{eq:xidt}
  \xi_{\bm{k}}(t)=\xi_{\bm{k}}^0 [1+\beta(t)].
\end{equation}
This corresponds to a change in the DOS, $N(t)=N_0/[1+\beta(t)]$ where $N_0$ is the equilibrium value. These equations show that by exciting the Raman phonon one can induce periodic oscillations in the DOS which will take the superconductor out of equilibrium.   

At this point it is important to emphasize the following: the thermodynamic BCS gap equation for a mechanism with cutoff frequency $\omega_{D}$,
\begin{equation}
 \Delta_0=2\omega_{D} e^{-\frac1{N_0\lambda}},
\end{equation}
suggests that a change in the DOS can be absorbed in a change in $\lambda$. Such an assumption has being done in the past. However, in the present formalism, changing $\lambda$ and changing the DOS through Eq.~(\ref{eq:xidt}) are  distinct drives that lead to distinct dynamics. Nevertheless, we find that the results are qualitatively similar.

\subsection{Phonon-assisted coupling constant driving }
\label{sec:phon-assist-coupl}


An interesting type of driving consist of phonons that can modulate the coupling constant. We will refer to this case as $\lambda$-driving. 
We propose that $\lambda$-driving can be realized in unconventional superconductors, in particular Fe-based superconductors. 
 We discus in more detail the case of FeSe where many key experiments are available and estimates of the magnitude of the relevant quantities can be done but we expect that a similar mechanism applies to other materials. 

A compilation 
of several experiments in different materials of the Fe-based superconductors family \cite{Mizuguchi2010,Johnston2010,Okabe2010,Huang2010,Imai2017} show that the critical temperature 
(and therefore the condensation energy) is very sensitive  to the anion height from the Fe layer, $z$. In particular, for FeSe \cite{Okabe2010}, the critical temperature changes from $T_\mathrm{c}=12$ $^{\circ}$K for $z=1.457$~\AA, to $T_\mathrm{c}=34$ $^{\circ}$K  for $z=1.427$ \AA.
This has being shown in experiments under pressure and it has being argued that the enhancement of $T_\mathrm{c}$ should originate on the crystal structure because the total carrier density of the FeSe layer does not change with pressure. 
Thus, this material is particularly appealing to modulate the order parameter with a time-dependent lattice distortion. As an order of magnitude estimate, the numbers quoted above amount to a rate of change of $T_\mathrm{c}$ with the anion height of $7$ $^{\circ}$K/pm.  

Similar conclusions can be drawn  from scanning tunneling microscopy experiments \cite{Song2012} which show that the superconducting gap decreases approaching twin-boundaries where the Se height is expected to increase.  
Also magnetic penetration depth measurements \cite{Hashimoto2012} suggest that the gap function is coupled to the pnictogen height to the point that even the symmetry of the order parameter can change with $z$. 

Theo\-re\-ti\-ca\-lly \cite{Yin2008,Mazin2008a,Yndurain2009,Sharma2009,Giovannetti2011,Sadovskii2016} one finds in these materials that the {\em magnetism} is very sensitive to the anion height $z$. This can be naturally explained if the system is close to a Stoner instability controlled by 
the latter.  Furthermore, for the magnetic mechanism expected for these  materials, the paring interaction is controlled by the magnetic susceptibility \cite{DeGennes1966mag,Maier2011}. Therefore, it is quite natural to attribute a
large fraction (if not all) of  the anion height sensitivity in $T_\mathrm{c}$ to  a modulation of the paring interaction via the magnetic susceptibility.

Another feature that makes Fe-based materials particularly suited for our propose is that practically in all materials there is an $A_{1g}$ phonon mode which involves the anion height coordinate. This mode can be launched in real 
time in a pump-probe experiment by a stimulated Raman process \cite{Merlin1997,Stevens2002,Lorenzana2013}. In the case of FeSe this was clearly shown recently by measuring the time-dependent anion height after a 
pump pulse \cite{Gerber2017}. Nicely, we can use these results to estimate the magnitude of the proposed effect.  
Exciting with a $1.5$ eV infrared light pulse and a fluence of about $0.46$ mJ/cm$^2$ Gerber {\it et al.} found that an amplitude of $\pm 0.25$ pm is achieved for the  $A_{1g}$ phonon that oscillates  at a frequency $\omega_d=5.3$ THz \cite{Gerber2017}. In the static limit, such a displacement would correspond to a  variation of $T_\mathrm{c}\sim \pm 2$ $^{\circ}$K which translates in a variation of about 10\% in the zero temperature equilibrium gap. 
Since the changes in the DOS within band theory are not of that order, it is natural to assume that such dramatic variations are due to phonon induced changes in the 
pairing interaction as explained above. Therefore, we consider that the coupling constant is a function of 
the anion height which can be manipulated with laser impulses.  
To estimate the change in $\lambda$ one can use the  BCS thermodynamic gap function to obtain, $\delta\lambda/\lambda=N_0\lambda \delta \Delta/\Delta$ where $N_0$ is the density of state at the Fermi level. Since $N_0\lambda<1$, the changes in $\lambda$ are smaller than 10\%. 
Below we will study modulations of $\lambda$ up to 10\% for illustrative proposes but our main results are robust and visible for much smaller values.  

The quoted fluence is close to the one that was used in Ref.~\cite{Mansart2013} ($0.3$ mJ/cm$^2$) 
to observe oscillations of the superconducting condensate in cuprates ($T_\mathrm{c}=40$ $^{\circ}$K). Therefore, it is also reasonable to assume that a similar 
fluence in a FeSe superconductor will not destroy the superconducting condensate.
Notice that the ISRS mechanism to launch the phonon does not require absorption. Therefore, the energy deposited in the sample could be further minimized by 
tuning the laser excitation energy to a transparent energy region of the material, allowing in principle to increase the amplitude of the oscillation without heating. 
In the following we will consider that,
\begin{equation}
  \label{eq:ldt}
 \lambda(t)=\lambda_0+\frac{d \lambda}{d u} u(t) =\lambda_0[1+\vartheta\left(t\right)\alpha\sin\left(\omega_d t\right)]\,,
\end{equation}
and  we will take the parameter   $\alpha\in\left[0,0.1\right]$.

In FeSe there is a dramatic enhancement of the equilibrium superconducting $T_\mathrm{c},$ as sample-thickness is reduced. $T_c$ changes from $8$ $^{\circ}$K in bulk to $77$ $^{\circ}$K in a monolayer grow on  SrTiO$_{3}$ which 
corresponds to a similar variation of the gap parameter.  Thus, by controlling the film thickness and assuming the phonon frequency does not change much, we have a wide range of the ratio $\omega_d/2\Delta_{0}\sim1-10$
which is experimentally accessibly and worth to explore in numerical simulations. 

\subsection{Impulsive Stimulated Raman Scattering (ISRS) Driving}
Electronic Raman active excitations of the condensate \cite{Mansart2013,Lorenzana2013} can be used to drive a superconductor through an ISRS process analogous to ISRS for phonons \cite{Merlin1997} or magnons \cite{Garrett1997}. 
Neglecting absorption at the pump-laser frequency, the total Hamiltonian for this process is $H=H_0+H_R$ with $H_0$ the equilibrium BCS Hamiltonian and $H_R$ the electronic Raman Hamiltonian \cite{Mansart2013,Lorenzana2013}, which is given by,   
\begin{eqnarray}
H_R&=& -2 \sum_{X,\bm{ k} } v_X(t) f_{\bm{k}}^X S_{\bm{k}}^{z},\nonumber\\
\label{eq:vx}
v_X(t)&=&-\frac12\bm{E}(t)\cdot\frac{\partial \bm{\chi}(\omega_L) }{\partial  N_X }\cdot\bm{E}(t)\,.
\end{eqnarray}
Here $\bm{E}(t)$ is the light electric field of the pump laser,
 ${\partial \bm{\chi}(\omega_L) }/{\partial  N_X }$ is the Raman tensor for electronic Raman scattering in symmetry $X$ and $\omega_L$ the laser carrier frequency. The description in terms of the electric field (instead of the vector potential as in next subsection) emphasizes the relation with the optical properties at the energy of the pump, $\hbar\omega_L$. 

Restricting to a tetragonal layered material, the most relevant symmetry functions are
\begin{eqnarray} 
  \label{eq:sqrharmonics}
  f_{\bm{k}}^{A_{1g}}&=&\frac12[\cos(k_x a)+\cos(k_y a)],  \nonumber\\
  f_{\bm{k}}^{B_{1g}}&=& \frac12[\cos(k_x a)-\cos(k_y a)], \nonumber\\
  f_{\bm{k}}^{B_{2g}}&=& \sin(k_x a)\sin(k_y a)\,.
\end{eqnarray}
Notice that we have not included $f_{\bm{ k}}^{A_{1g}}=1$ as it leads to driving by the total number operator which is irrelevant for the reasons explained in Sec.~\ref{sec:model}.
Eq.~(\ref{eq:vx}) shows that in this case the laser electric
field acts as a time-dependent potential acting on charge excitations
with different symmetries. Usually pulses of about $50$ fs can be produced which modulate an IR or visible laser. Mansart {\it et al.} have shown \cite{Mansart2013} that one such pulse 
induces  a fluctuation of the condensate at a frequency close to $2\Delta$ by an ISRS process. Also here it is possible to adjust $\omega_L$ to a window of low absorption to minimize heating. 

The charge fluctuation after one pulse decays very rapidly because of dephasing of excitations with different frequency. However, as mentioned in 
\cite{Lorenzana2013}, one can excite the material with a periodic sequence of pulses so that the excitations that match the periodicity of the pump are reinforced and other excitations are suppressed. It is precisely the dynamics of 
this kind of excitation that is described in detail below.  One advantage of this method is that it is very easy to detect the response of the condensate through the modification of the optical properties at optical or other 
frequencies \cite{Mansart2013,Lorenzana2013}. 

Selection rules for ISRS are similar to the one of spontaneous Raman scattering except that in the former 
$v_X$ involves the same polarization of the electric field at both sides of the Raman tensor in Eq.~(\ref{eq:vx}) while in the latter two different fields appear related to incoming and outgoing photons. Therefore, in the notation 
of Table I of Ref.~\cite{Devereaux2007},
polarizations $xx$, $yy$ and $x'x'$ are accessible in ISRS while $xy$ is not.  
These ``parallel'' polarizations excite the $A_{1g}$ symmetry modes plus other modes.
By symmetry, and using the same arguments as before, the $A_{1g}$ part of the drive can be taken to the DOS-driving form of Eq.~(\ref{eq:xidt}) plus sub-leading terms 
with more complicated structure along the Fermi surface. This is obvious in the case of a lattice model with only nearest-neighbor hopping 
since the $A_{1g}$ symmetry function is proportional to the dispersion relation. For reasons of simplicity, in our simulations we concentrate
on modes that preserve the symmetry of the lattice but we expect that 
for modes of lower symmetry very similar physics arises.

\subsection{Direct THz drive}
Matsunaga {\it et al.} have shown \cite{Matsunaga2013,Matsunaga2014} that THz radiation pulses can produce oscillations of
a superconducting condensate. They interpreted this result as due to the coupling of the Higgs amplitude mode to the THz electric field. However, this interpretation has being disputed by Cea, Castellani and Benfatto \cite{Cea2016b}
who argued that the response is dominated by charge fluctuations similar to the ones of the transient Raman experiment of Mansart {\it et al.} \cite{Mansart2013}. Notwithstanding,  this kind of drive is very interesting since it is possible to apply a THz radiation with a frequency smaller than the gap in such a way that, to leading order, there is no direct excitation of  quasi-particles and one expects much less heating than with a ISRS drive.

We introduce the coupling with the electromagnetic field through the Peierls substitution for carriers with charge $q$ ($ =-|e|$ for electrons) in the underlying lattice model,
\begin{equation}
c_{i+\bm{r}}^\dag c_{i}\rightarrow c_{i+\bm{r}}^\dag c_{i}e^{-i \bm{a}\cdot\bm{ r}},
\end{equation}
where $\bm{ a}\equiv q \bm{  A}/(\hbar c)$ and we assume that  the vector potential $ \bm{A}$ is uniform on the scale of the sample giving rise to the electric field $\bm{E}=-\dot{\bm{A}}/c$.  

Expanding up to second order in the vector potential we have that the Hamiltonian becomes $H=H_0+H_{T}$ with 
  \begin{equation}
    \label{eq:ha}
H_T=-2\sum_{\bm{k}} \left(
\frac{\partial\varepsilon_{\bm{k}}^0}{\partial\bm{k}}\cdot\bm{ a} (t)+
\frac12\frac{\partial^2\varepsilon_{\bm{k}}^0}{\partial\bm{k}_\mu\partial\bm{k}_\nu} a_\mu(t) a_\nu(t)
\right) S_{\bm{k}}^{z},
  \end{equation}
where the sum over repeated indexes is implicit. 
The first term inside the parenthesis represents the coupling of the vector potential to the paramagnetic part of the current.  
Consider $\bm{A}(t)=\bar{\bm{A}}\, \theta(t) \sin(\omega_{T} t)$.
For $\omega_{T}<2\Delta$ and not very strong disorder, there is no absorption from the paramagnetic part (the real part of the optical conductivity is zero) and there is not transfer of energy between the drive and the system. 
Since we are interested in processes where the drive and the system exchange energy, we consider only the second term. 
This behaves as a drive with a time dependence $ \sim\cos(2\omega_{T} t)$. 
If we now consider a system with the same symmetries of the previous subsection and  the $x'x'$ polarization so that $\bar A_x=\bar A_y$ we reach again a momentum dependence with the same symmetry as the dispersion relation (plus higher order corrections). 
In analogy with the ISRS case, this  can be taken to the DOS-driving form of Eq.~(\ref{eq:xidt}). 
Notice that due to the similarity with the ISRS case it is quite tempting to excite in the THz to avoid heating but to use a probe at optical energies where it is easy to achieve high resolution in time. 

\subsection{Drive in ultra-cold atoms}
The manipulation of the Hamiltonian parameters in ultra-cold atoms is very well known \cite{Bloch2008} and therefore we mention it here very briefly. 
Both $\lambda$-driving and DOS-driving can be  technically achieved in  ultra-cold atoms. 

 The interaction between fermions can be controlled through a 
time-dependent magnetic field that modulates a 
Feshbach resonance \cite{Chin2010} or through optical control which is much faster \cite{Clark2015}. Very recently,  Ref.~\cite{Behrle2018} has introduced a novel way to modulate the order parameter which appears well suited for our propose.  This method involves generating Rabi oscillations between one of the two fermionic states participating in pairing and a third state.  

To modulate the DOS one can consider fermions moving in an optical lattice \cite{Bloch2008}. It is possible, then, to modulate the depth of the potential well in time as has already been done for bosonic systems \cite{Stoferle2004,Haller2010,Endres2012}. 
Such drive modulates the DOS via the change in the hopping integral.

\section{Linear Response}
As mentioned above, even though the DOS-driving and $\lambda$-driving are different, the main physical results are very similar. Thus, in the remainder of this work
we analyze the $\lambda$-driving in further detail.

As a warmup exercise we compute the linear response of the superconductor to an harmonic $\lambda$-drive with adiabatic switching.
In the following, we denote the corrections that are linear in the perturbation by a superscript ``$1$'', that is $\bm{b}_{\bm{k}}=\bm{b}_{\bm{k}}^0+\bm{b}_{\bm{k}}^1 \exp[i(\omega -i\delta)t]$,
$\lambda=\lambda_0+\lambda_1\exp[i(\omega-i\delta) t]$ with $\lambda_{1}=\alpha \lambda_0$ and $\bm{S}_{\bm{k}}(t)=\bm{S}_{\bm{k}}^{0}+\bm{S}_{\bm{k}}^{1} \exp[i(\omega-i\delta) t]$ where $\delta$ is an infinitesimal positive quantity. 
The linearized equation of motion becomes
\begin{equation}
  \label{eq:leom}
 i\omega  \bm{S}_{\bm{k}}^1=-\bm{b}_{\bm{k}}^1\times   \bm{S}_{\bm{k}}^0-\bm{b}_{\bm{k}}^0 \times   \bm{S}_{\bm{k}}^1\,,
\end{equation}
with solution
\begin{equation}
 \label{eq:sk1}
\bm{S}_{\bm{k}}^1=
 \frac{{b^{x,1}_{\bm{k}}} \xi_{\bm{k}}}{2\sqrt{{\Delta_0}^2+\xi_{\bm{k}} ^2}\left((\omega-i\delta)^2-4 \left({\Delta_0}^2+\xi_{\bm{k}} ^2\right)\right) } 
\left(\begin{array}{c}
   -2\xi_{\bm{k}}  \\ 
-i  \omega   \\ 
 2{\Delta_0} 
\end{array}\right)\,.
\end{equation}
Notice that since we are considering an s-wave superconductor there is no 
momentum dependence of the $x-$component of pseudomagnetic field ($b^{x,1}_{\bm{k}}\equiv b^{x,1}$). Therefore, assuming a particle hole symmetric DOS the only 
non-zero component after summing over ${\bm{k}}$ is   
\begin{equation}
 \label{eq:s1t}
S_t^{x,1}=\chi_{\Delta,\Delta}^0(\omega)  b^{x,1}\,,
\end{equation}
where the quantity on the left side was defined below Eq.~(\ref{eq:del2}) and  
we introduced the bare susceptibility,
\begin{equation} 
\label{eq:chisum}
\chi_{\Delta,\Delta}^0(\omega)= -\sum_{\bm{k}}\frac{{\xi_{\bm{k}}^2} }{\sqrt{{\Delta_0}^2+\xi_{\bm{k}} ^2}\left((\omega-i\delta)^2-4 \left({\Delta_0}^2+\xi_{\bm{k}} ^2\right)\right)}\,.
\end{equation}
The imaginary part of Eq.~(\ref{eq:chisum}) is given by
\begin{equation}
\label{eq:imag}
 \mathrm{Im}\left(\chi_{\Delta,\Delta}^{0}(\omega)\right)=-\sgn(\omega)\frac{\pi\rho}{4}\sqrt{1-\left(\frac{2\Delta_{0}}{\omega}\right)^{2}},
\end{equation}
for $2\Delta_{0}<|\omega|<\omega_{D}$ and zero otherwise, while the real part can be obtained  from the Kramers-Kroning relation
\begin{equation}  
\label{eq:real}
\mathrm{Re}\left(\chi_{\Delta,\Delta}^{0}(\omega)\right)=\frac{1}{\pi}\mathcal{P}\int_{-\infty}^{\infty}\frac{\mathrm{Im}\left(\chi_{\Delta,\Delta}^{0}(\omega')\right)}{\omega'-\omega}d\omega'.
\end{equation}
Furthermore, we can write
\begin{equation}
b^{x,1}(t)=2\Delta_1(t)=2\left(\lambda_1(t) S_t^{x,0}+\lambda_0 S_t^{x,1}(t)\right),
\end{equation}
and by using Eq.~(\ref{eq:s1t}) we obtain the pseudo-spins response for a positive frequency $\omega$ 
\begin{equation}
  \label{eq:chi}
S_t^{x,1}(t)=\alpha\chi_{\Delta,\Delta}(\omega) S_t^{x,0} e^{i\omega t},
\end{equation}
where we have defined
\begin{equation}
\chi_{\Delta,\Delta}(\omega)\equiv \frac{2 \lambda_0\chi_{\Delta,\Delta}^0(\omega)}{1-2\lambda_0 \chi_{\Delta,\Delta}^0(\omega)}.
\label{eq:chi1}
\end{equation}
Summing the response for $\pm\omega$ and taking the real part, we get  
that the correction to the gap amplitude $\Delta_1$ induced by a $\cos(\omega t)$ drive is
\begin{equation}
\label{eq:delta1}
\Delta_{1}=\alpha\left|1+\chi_{\Delta,\Delta}(\omega)\right|\Delta_{0}\,,
\end{equation}
where we have used that $\Delta_{0}=\lambda_{0} S_t^{x,0}.$

\begin{figure}[tb]
\includegraphics[width=0.4\textwidth]{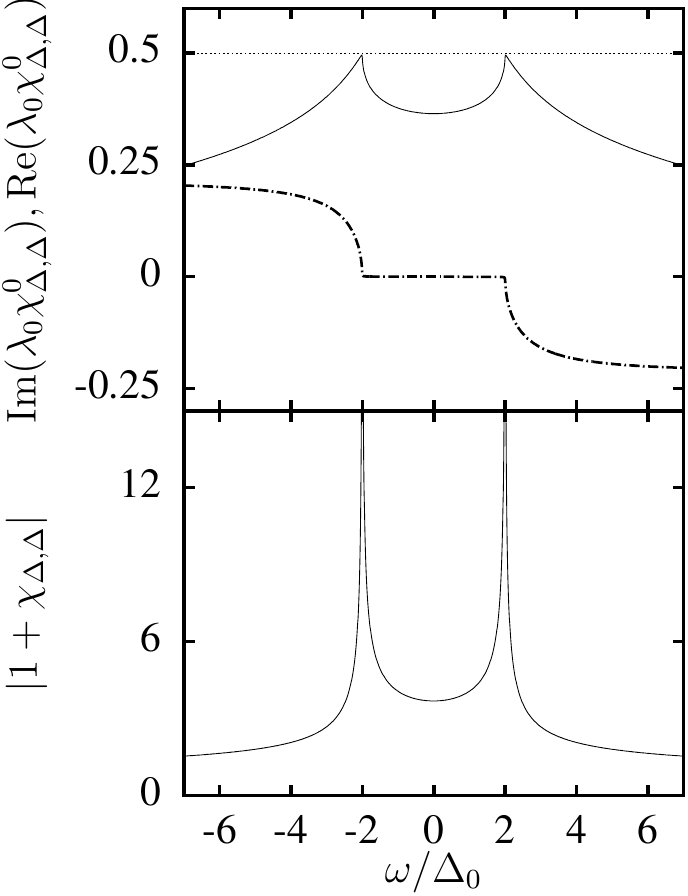}
\caption{ Top panel: the real and imaginary part of susceptibility Eq.~(\ref{eq:real}) and Eq.~(\ref{eq:imag}) times $\lambda_0$ is shown as a continuous line and dot dashed line respectively. The horizontal dashed line shows that the equation $1-2\lambda_0 \chi_{\Delta,\Delta}^0(\omega)=0$ is satisfied at 
  $\omega=2\Delta_0$. Bottom panel: the response determining the amplitude of the order parameter oscillations excited by a $\lambda$-drive. 
 We use a cutoff frequency $\omega_{D}=20\Delta_{0}$.} 
\label{fig:chi}
\end{figure}

In Fig.~\ref{fig:chi} we show the imaginary and real part of the function $\chi_{\Delta,\Delta}^0(\omega)$. 
It is easy to show \cite{Kulik1981} that the real part of the denominator of
Eq.~(\ref{eq:chi1}) has a pole at $\omega=2\Delta_0$ as it is apparent in the figure. This produces the well know Higgs-like resonance in $\chi_{\Delta,\Delta}$ which 
has been emphasized in the context of Raman scattering \cite{Cea2014}, THz drive \cite{Matsunaga2014,Tsuji2015} and driven cold-atoms\cite{Behrle2018}.
The figure also shows 
$\left|1+\chi_{\Delta,\Delta}(\omega)\right|$ which determines the amplitude of the oscillation under drive and presents a resonant behavior. We will show that 
going beyond linear response even away from the Higgs resonance interesting effects arise. 

The imaginary part of $\chi_{\Delta,\Delta}$ describes transfer of energy from the drive to the system. The validity of the 
equations requires a small drive but also times that are not too long. Indeed, at long times one has to describe the fate of this energy. In a closed system
(as an ultra-cold atom system is in first approximation)  
one may think that one would reach thermalization at infinite temperatures at very long times. However, as we will shown below, going beyond linear response 
this does not occur in the present system, a fact that can be attributed to the integrability of the model \cite{Lazarides2014}. In an open system in contact with a thermal bath one expects that the energy will be transferred to 
the bath, usually the lattice phonons in a solid state state superconductor. 
However, at low temperature these times can be very long \cite{dynes1978}
leaving a large time window were the system effectively behaves as if were closed.


\section{Non-Linear response and Rabi-Higgs modes}
We now explore the behavior of the system when the drive acts for long times and/or the drive amplitude is not small. 
\subsection{Numerical Results}

We shall show the numerical calculation of $\Delta(t)$ beyond linear response. In our computations, for $t\leq0$ the system is in equilibrium. The superconducting gap is set to $\Delta_{0}$
and assumed to be real.  
The zero-temperature pseudo-spin texture is given by Eq.~(\ref{eq:sequilib}).
At $t>0$ the drive switches on according to Eq.~(\ref{eq:xidt}) or (\ref{eq:ldt}) and the pseudo-spins evolve according to Eq.~(\ref{eq:eom}),
which in turn will change the gap $\Delta\left(t\right)$ and the local
fields $\bm{b}_{\bm{k}}\left(t\right)$. 
We take a set of $N$ pseudo-spins uniformly spaced in $\xi_{\bm{k}}$ within a band of width $\mathcal{W}=40\Delta_{0}=2\omega_{D}$ which equals twice the cutoff frequency ($\omega_{D}$, for conventional superconductors). Fourth-order Runge-Kutta method was used to numerically integrate the spin 
equations of motion with $N=4\times 10^{4}$. If the initial order parameter is real (a gauge choice), then the symmetry of the problem dictates that it remains real 
at all times in both the case of $\lambda$- and DOS-driving. Hence, we set Eq.~(\ref{eq:del2}) to be zero.  

We assume a driving amplitude that is constant in time. In the case of phonon assisted driving, this may appear unrealistic as real phonons will have damping. More so in the presence of the superconductor where energy will be transferred from the phonon to the superconductor. However, in experiments this can be compensated by periodically applying pulses with a periodicity which is a multiple of the phonon period in order to restore the lattice oscillation 
to the original amplitude. Obviously, in the case of ultra-cold atoms, electronic ISRS or THz driving this problem does not arise.

In the following we present the results for the $\lambda$-driving of Eq.~(\ref{eq:ldt}) in detail. We use parameters adequate for the phonon assisted version in FeSe materials (Sec.~\ref{sec:phon-assist-coupl}). 
Qualitatively equivalent result were found with the DOS-driving and with other parameters.

\begin{figure}[tb]
\raggedright{}\includegraphics[width=0.47\textwidth]{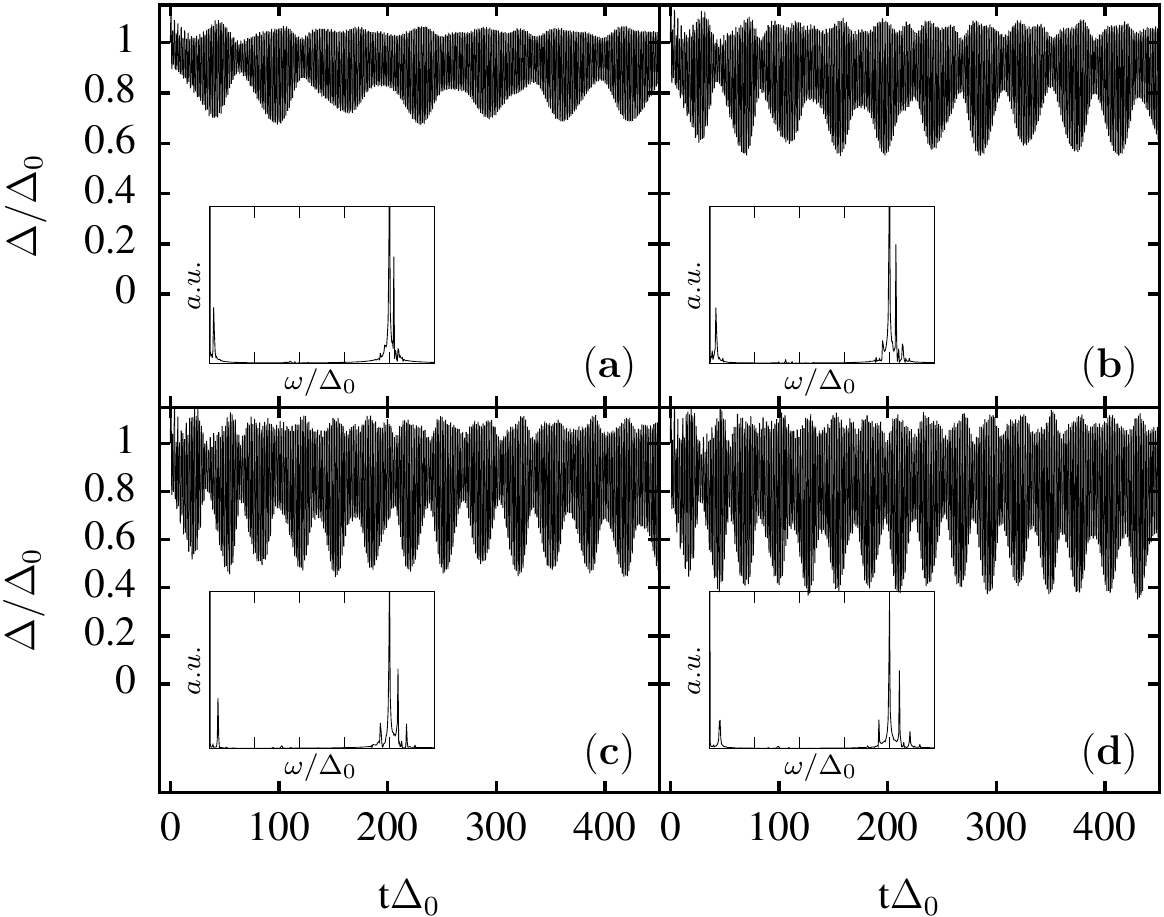}
\caption{Time dependence of superconducting order parameter for $\alpha=0.04$ (a), $\alpha=0.06$ (b), $\alpha=0.08$ (c) and $\alpha=0.1$ (d). Fast Fourier transform is shown in the insets. Two fundamental frequency 
appear in the spectrum corresponding to the drive frequency $\omega_d=4\Delta_{0}$ and a small frequency $\omega_{R}.$ Satellite peaks at $\omega_d\pm n \omega_{R}$ with $n=1,2,3$ are observed. }
\label{fig:tdd} 
\end{figure}

The dynamics of $\Delta\left(t\right)$ is shown in Fig.~\ref{fig:tdd} for $\omega_d=4\Delta_0$ and several values of $\alpha$. The first feature to notice is that despite the drive can give energy to the superconductor indefinitely, the average order parameter decrease from the equilibrium value but it is not totally suppressed, i.e. the system is not driven to infinite temperature. 

In Figure~\ref{fig:tdd} (and Fig.~\ref{fig:deltavsomega} below), the drive frequency correspond to an  oscillation that is too fast to be resolved 
on the scale of the figure and leads to the filled black regions. The system essentially synchronizes with the drive. 
Both features, synchronization and absence of heating are  expected \cite{Lazarides2014} 
 for an integrable system as the present one. 
On top of that, the amplitude of the gap shows slow oscillations.
This new non-linear low-frequency mode is our main result and will be dubbed 
Rabi-Higgs mode. 

The period of the Rabi-Higgs mode
decreases with increasing $\alpha$ as will be discussed in detail below. 
It is important to realize that these driven modes are very different from the spontaneous Higgs modes reported before \cite{Barankov2004,Barankov2006a,Yuzbashyan2006} 
which have a different frequency and do not show a similar sensitivity to the drive.

In order to gain more information on the oscillations we perform a fast Fourier transform analysis. This is shown in the insets of Fig.~\ref{fig:tdd}. We find that for all studied $\alpha$ there are 
two fundamental frequencies in the dynamical response of $\Delta\left(t\right)$. The first is the drive frequency
$\omega_d$ and the second one corresponds to the Rabi-Higgs frequency $\omega_{R}$ with period $\tau=2\pi/\omega_{R}$. Moreover, satellite peaks can be observed at $\omega_d\pm n \omega_{R}$, with $n$ a small integer.  

\begin{figure}[tb]
\raggedright{}\includegraphics[width=0.47\textwidth]{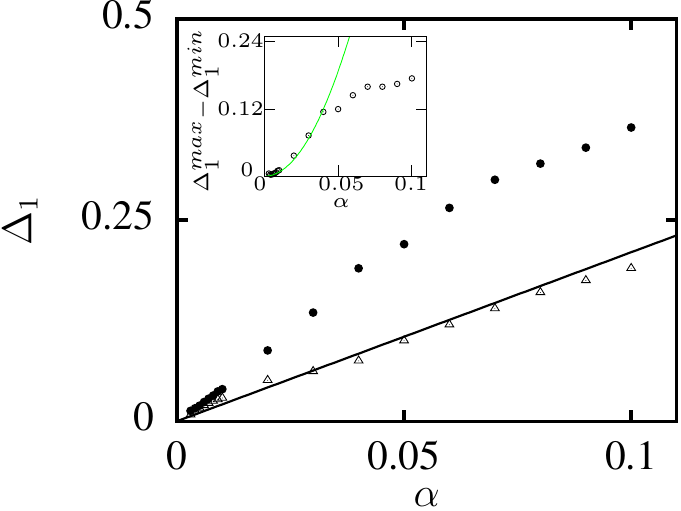}
\caption{(Color online) Amplitude of the oscillation in the order parameter as a function of the strength of the drive. For the definition of the minimum (triangles)  and maximum (filled circles) amplitudes see text. The solid line is the linear response result of 
Eq.~(\ref{eq:delta1}) with $\omega=\omega_d$. The inset show the difference between 
the two amplitudes for small $\alpha$. The empty dots are obtained from numerical result while the green line is a quadratic fitting.}
\label{fig:delta1} 
\end{figure}

We now discuss the amplitude of the order parameter oscillations at frequency $\omega_d$. Because of the Rabi-Higgs mode, the amplitude of the oscillations 
(around some average value) is not constant leading to the variable width 
of the black regions in Fig.~\ref{fig:tdd}.
In order to roughly take into account this effect,  we define an effective time-dependent amplitude in each 
interval $[t,t+2\pi/\omega_d]$. The minimum and maximum of such time-dependent amplitude in a long steady-state part of the dynamics are denoted as $\Delta^{\mathrm{min}}_{1}$ and $\Delta^{\mathrm{\mathrm{max}}}_{1}$, respectively. In practice, they are the 
minimum and maximum widths of the black regions in Fig.~\ref{fig:tdd} at long times. Fig.~\ref{fig:delta1} shows the two amplitudes as a function of the strength of the drive $\alpha$. The solid line is the linear response result of 
Eq.~(\ref{eq:delta1}) with $\omega=\omega_d$. We see that the latter predicts the correct magnitude of  $\Delta_1$ and it is quite close to $\Delta_1^{\mathrm{min}}$, i.e. linear response fixes the scale of the amplitude of the oscillations at the 
drive frequency even far form its strict range of strict validity (small $\alpha$ and short times). Linear response can not explain the difference between $\Delta_1^{\mathrm{min}}$ and $\Delta_1^{\mathrm{max}}$ which is an exquisitely 
non-linear effect.
Indeed such difference is associated with the appearance of a new frequency which is clearly an effect beyond linear response. This can be also seen from the inset of Fig.~\ref{fig:delta1} where it is clear from the numerical results 
that for small $\alpha$ the difference between the two amplitudes is approximately quadratic in $\alpha$.  There is a small mismatch of the slope at small $\alpha$ but this is not surprising as even in that regime we are not strictly in  the
conditions of validity of linear response  since the switching was not adiabatic and the time of the measurement was not small.  
What is important for our propose is that the overall scale is well predicted.
As a further check that the scale of the amplitudes are determined by the linear responses susceptibility we change the drive frequency towards the 
Higgs resonance of Fig.~\ref{fig:chi}. As expected, as $\omega_d\rightarrow2\Delta_0$ the amplitude increases as is shown in the Fig.~\ref{fig:deltavsomega}.

\begin{figure}[b]
\raggedright{}\includegraphics[width=0.47\textwidth]{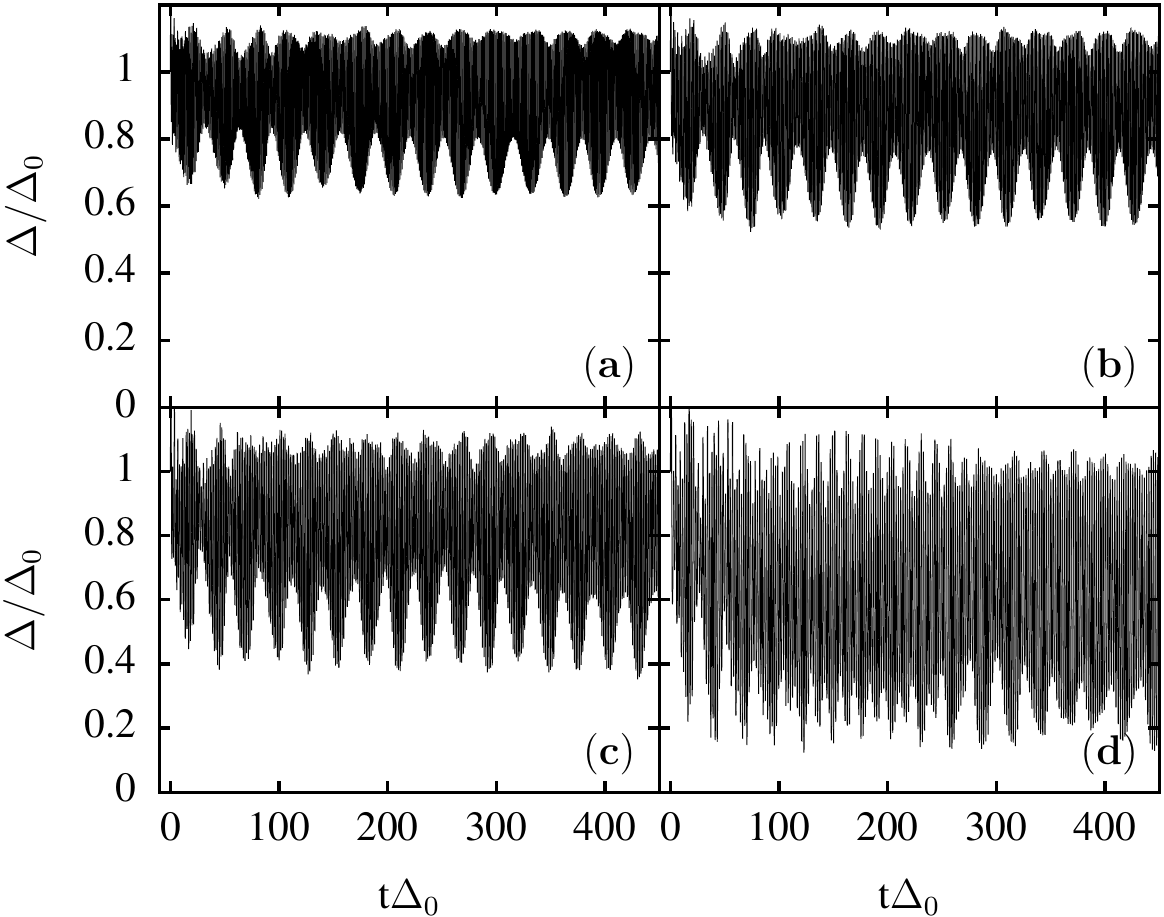}
\caption{Time dependence of superconducting gap for $\omega_d=6\Delta_{0}$ (a), $\omega_d=5\Delta_{0}$ (b), $\omega_d=4\Delta_{0}$ (c) and $\omega_d=3\Delta_{0}$ (d) with $\alpha=0.1$. 
The amplitude of the order parameter increase as $\omega_d\rightarrow2\Delta_0$.}
\label{fig:deltavsomega} 
\end{figure}

In the remainder of the paper, we present simulations only for a drive frequency $\omega_d=4\Delta_{0}$ but qualitatively similar results 
were found for $\omega_d=6\Delta_{0}$ and $\omega_d=8\Delta_{0}$.

\subsection{NMR analogy}
To the best of our knowledge the drive induced Rabi-Higgs mode
has not been reported before.  It originates in a resonant phenomenon for pseudo-spins analogous to Rabi oscillations in 
NMR experiments for usual spins as we shall demonstrate below.
 
In the presence of a static magnetic field $B_{0}$ any spin or magnetic moment precesses with the Larmor frequency $\omega_{L}$ which is proportional to $B_{0}$. If a small alternating magnetic field of amplitude $B_{1}$ is 
applied with a frequency $\omega=\omega_{L}$ in a plane perpendicular to $B_{0}$, the spin experience Rabi oscillations with a new low frequency which is proportional to $B_{1}$ \cite{Slichter1990} 
\begin{equation}
  \label{eq:rabi}
\omega_R=\gamma B_{1}\,,  
\end{equation}
where $\gamma$ is the gyromagnetic ratio.
\begin{figure}[tb]
\includegraphics[width=0.3\textwidth]{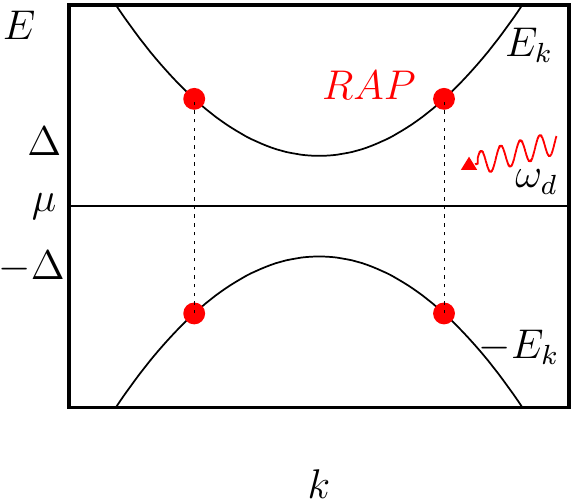}
\caption{(Color online) Schematics of the proposed experiment. A drive at frequency $\omega_d$ 
enters into resonance with a set of Bogoliubov quasiparticles having energy $\pm E_{\bm{ k}}= \pm \sqrt{\xi_{\bm{k}}^{2}+\Delta_{0}^{2}}$. The RAP set is indicated by the red dots in this one-dimensional cut and
corresponds to a line for the case of a two-dimensional superconductor.
} 
\label{fig:exp}
\end{figure}

\begin{figure}[tb]
\includegraphics[width=0.47\textwidth]{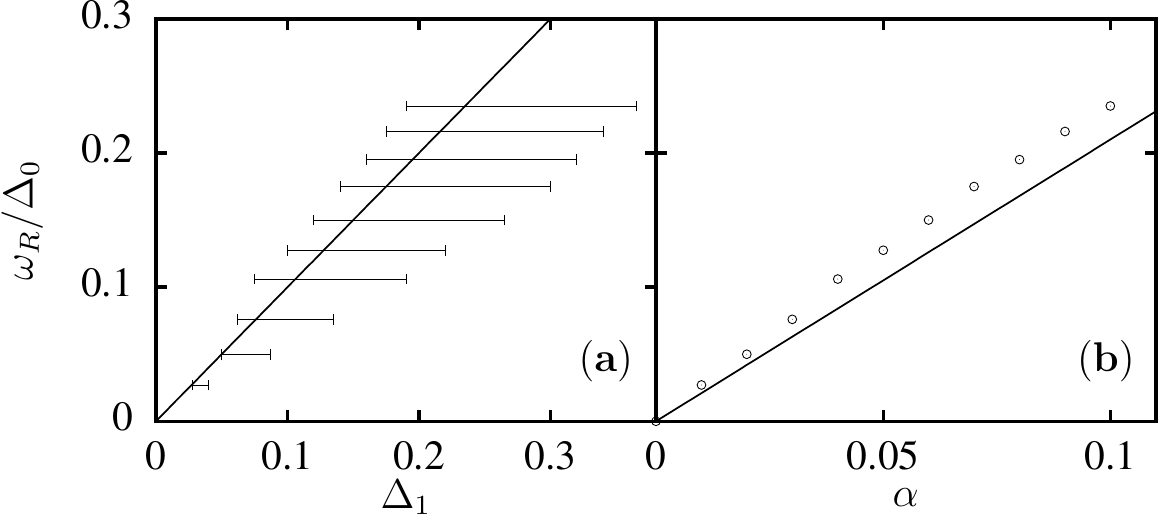}
\caption{(a) Dependence of low-energy mode frequency $\omega_{R}$ on $\Delta_1$ both extracted from the numerical simulations. The line corresponds to the prediction of Eq.~(\ref{eq:w0}) assuming a constant amplitude  $\Delta_1$ and making the approximation
$\bar \Delta=\Delta_0$. The error bars indicate the range $\Delta^{\mathrm{min}}_{1}<\Delta_1 <\Delta^{\mathrm{max}}_{1}$ (see  text). (b) $\omega_{R}$ as a function of 
$\alpha.$ The empty dots represent $\omega_{R}$ extracted from simulations while the solid line is the estimated value after Eq.~(\ref{eq:w0}) where $\Delta_1$ depends on $\alpha$ according to Eq.~(\ref{eq:delta1}). The final expression is display in Conclusions as Eq.~(\ref{eq:wr}). 
}
\label{fig:w0dd1}
\end{figure}

In the case of the pseudo-spins the role of the static magnetic field is played by the mean-field $\bm{b}_{\bm{k}}$ and the Larmor frequency is,  $\omega_{L}=2\sqrt{\xi_{\bm{k}}^{2}+\Delta_{0}^{2}}$. This is nothing but the energy needed to create two Bogoliubov quasiparticles as shown schematically in Fig.~\ref{fig:exp}.
There is a family of pseudo-spins, labeled 
$\bm{S}_{\bm{k}_{r}}$,  for which the resonance condition
\begin{equation}
  \label{eq:res}
 \omega_d=2\sqrt{\xi_{\bm{k}_{r}}^{2}+\Delta_{0}^{2}},   
\end{equation}
is satisfied. We will refer to this family as resonant Anderson pseudo-spins (RAP). 
The pseudo-magnetic field acting on the RAP set is
\begin{equation}
 \bm{b}_{\bm{k}_r}\left(t\right)=2\left(\Delta(t),0,\xi_{\bm{k}_r}\right).
\end{equation}
As a first approximation we schematize the time dependence of the field by
\begin{equation}
 \Delta(t)=\bar\Delta+\Delta_1 \cos(\omega_dt),
\end{equation}
where $\bar\Delta$ is the average value of the order parameter in the steady state (similar but not necessarily equal to $\Delta_0$) and $\Delta_1$ is a constant oscillation amplitude at the frequency of the drive and whose magnitude can be approximated by linear response.
 We now decompose the time dependent part of the pseudo-magnetic field (which is directed along $x$) in a component parallel to the time independent pseudo-field and another component perpendicular to it. The perpendicular component is
\begin{equation}
 b_{\bm{k}_r}^{\perp}(t)=2 \Delta_1 \cos(\omega_dt) \sqrt{1-\left(\frac{2\bar \Delta
}{\omega_d}\right)^2}\,.
\end{equation}
According to Eq.~(\ref{eq:rabi}), this time-dependent magnetic field produces Rabi oscillations of the RAP with frequency
\begin{equation}
  \label{eq:w0}
\omega_R= \Delta_1 \sqrt{1-\left(\frac{2\bar \Delta
}{\omega_d}\right)^2}, 
\end{equation}
proportional to $\Delta_1$.
To check that $\omega_{R}$ arises from this underling mechanism we extracted, for each value of $\alpha$, $\Delta_1$ and $\bar\Delta$ from the full numerical 
solution for $\omega_R$ and compared it with Eq.~(\ref{eq:w0}). Fig.~\ref{fig:w0dd1} (a) shows the plot of $\omega_R$ as a function of $\Delta_1$.
The minimum ($\Delta^{\mathrm{min}}_{1}$) and maximum ($\Delta^{\mathrm{max}}_{1}$) of the order parameter are represented with an horizontal error bar. 
There is good agreement with the simulations with an effective fixed $\Delta_1$ within its physical range of variation.  Figure~\ref{fig:w0dd1}(b) shows
the frequency of the Rabi-Higgs mode as a function of the drive strength. 
The line is  Eq.~(\ref{eq:w0})  
where the value of $\Delta_1$ chosen is the one predicted by linear response, Eq.~(\ref{eq:delta1}). This approximation 
appears to be surprisingly good as seen in the figure.

\begin{figure}[tb]
\includegraphics[width=0.4\textwidth]{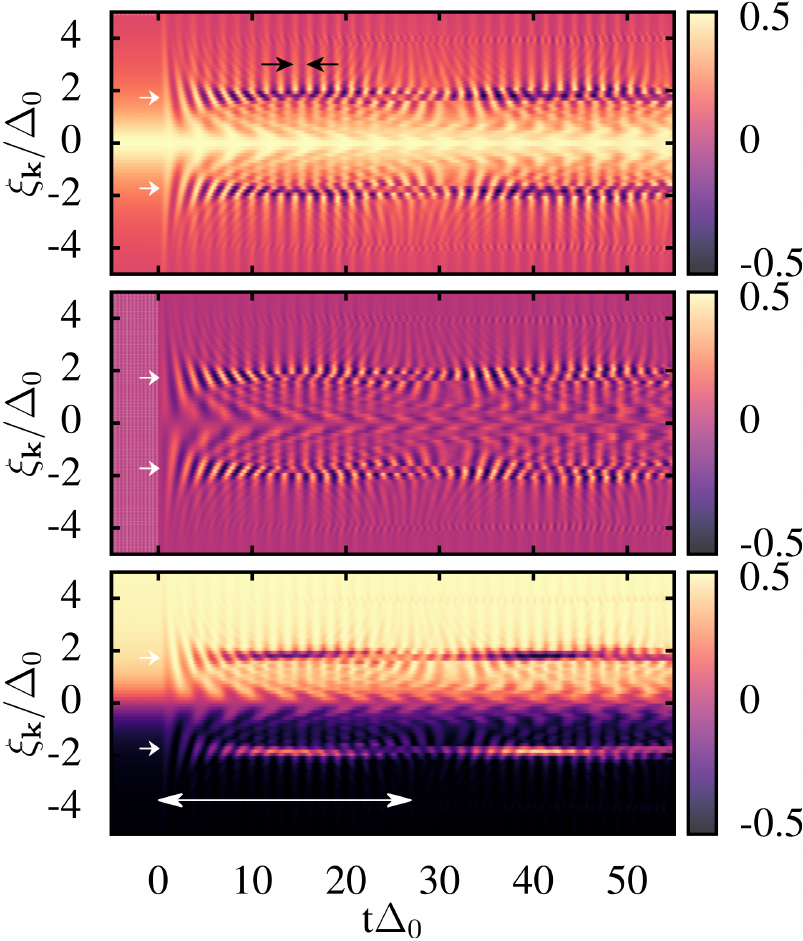}
\caption{(Color online) The pseudo-spins evolution as a function of time with $\alpha=0.1$. Top, middle and bottom panel correspond to $S_{\bm{k}}^{x}$,
$S_{\bm{k}}^{y}$ and $S_{\bm{k}}^{z}$ respectively. The equilibrium texture of pseudo-spins is shown at negative value of time (see Eq.~(\ref{eq:sequilib})). 
RAP with energy $\pm\sqrt{3}\Delta_{0}$ has been indicated by white arrows. The period associated with drive frequency is marked with black arrows in the top panel. On the other hand, a period $\tau\Delta_0\simeq27$ which appears in the 
RAP dynamics is pointed by white double arrow in the bottom panel. This period match with the period $\tau$ associated with the Rabi-Higgs mode of $\Delta\left(t\right).$} \label{fig:spins}
\end{figure}
\subsection{Physical consequences of Rabi-Higgs modes: population inversion}
We now analyze how the Rabi-Higgs mode affects different observables. 
In Fig.~\ref{fig:spins} the time dependence of the pseudo-spin texture is shown.
Pseudo-spins are labeled by the quasi-particle energy and shown as a function of time with negative times representing the equilibrium situation.  
Colors encode the projection of the pseudo-spins in the different directions. 
Notice that $\langle S_{\bm{k}}^{y} \rangle$ fluctuations, which determine the imaginary part of the order parameter [c.f. Eq.~(\ref{eq:del2})], are odd so they 
cancel when summed over $\bm{k}$. Therefore, only charge fluctuations and amplitude fluctuations enter into play in the integrated quantities (top and bottom panels,
respectively). 

 The drive frequency is visible in the dynamical response of all projections.
Its period $\mathcal{T}=2\pi/\omega_d$ corresponds to the vertical features indicated by black arrows in the top panel of Fig.~\ref{fig:spins}.
We find that most of the pseudo-spins only precesses, with small fluctuations, around their equilibrium values. In contrast, RAP laying at  $\xi_{\bm{k}_{r}}$ and indicated by the white horizontal arrows, respond strongly to the drive and 
visit the whole Bloch sphere in 
the Rabi period $\tau$, which is indicated by the white horizontal double arrow
in the bottom panel of Fig.~\ref{fig:spins}---it coincides with the period of the Rabi-Higgs mode. 
As in the analogous NMR experiment, RAP perform slow oscillations from the equilibrium position to the antipode in the Bloch sphere and rapid rotations around the axis joining these points. Such direction corresponds approximately to the 
equilibrium field  $\bm{b}_{\bm{k}_r}^0=2\Delta_{0}\,\hat{\bm{x}}+2\xi_{\bm{k}_r}\,\hat{\bm{z}}$
and is a function of $\omega_d$ through the resonant condition Eq.~(\ref{eq:res}).

Notice that the observed phenomena requires that a macroscopic number of pseudo-spins synchronize due to interactions. Thus the RAP set is not limited to the only ones strictly satisfying the resonance condition (\ref{eq:res}) but
there is a distribution in quasi-particle energies with a finite width around a central value which participate in the process.

\begin{figure}[tb]
\includegraphics[width=0.33\textwidth]{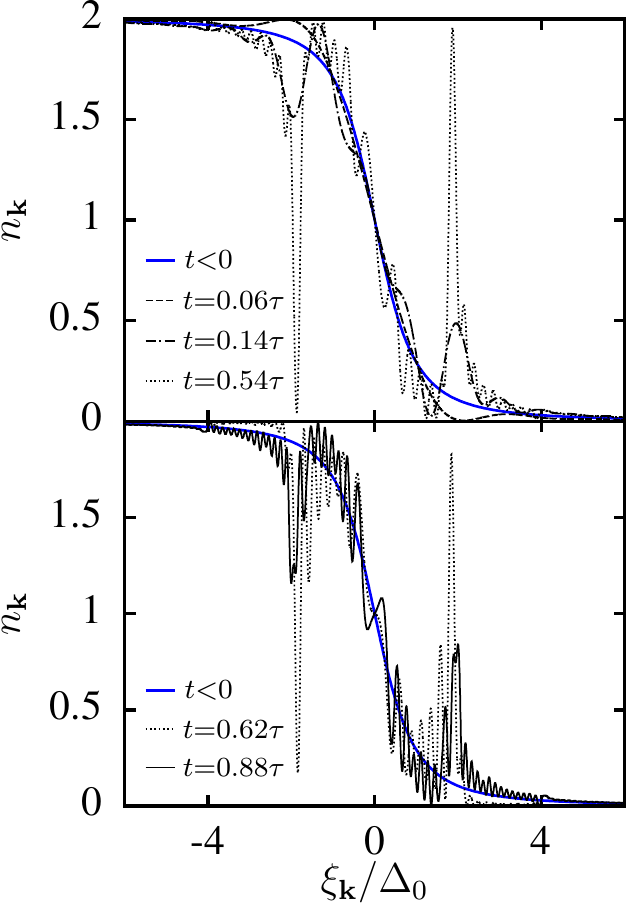}
\caption{(Color online) Occupation values $n_{\bm{k}}$ at different times for $\alpha=0.1$ ($\tau\Delta_0\simeq27$). Due to RAP show a periodic population inversion,
we obtain a strong $n_{\bm{k}}$ fluctuation for these resonant states. The occupation values at equilibrium are indicated by continuous blue line and 
denoted as $t<0.$} \label{fig:spins_cut}
\end{figure}

A major physical consequence of the Rabi-Higgs mode is a population inversion of the RAP occurring with periodicity $\tau$.
Fig.~\ref{fig:spins_cut} shows this phenomena in the occupation values $n_{\bm{k}}$ obtained through cuts of the lower panel of Fig.~\ref{fig:spins} at different times using 
the relation $n_{\bm{k}}=n_{\bm{k}\uparrow}+n_{-\bm{k}\downarrow}=1-2\langle S_{\bm{k}}^{z}\rangle.$ 
One clearly sees that the occupation values $n_{\bm{k}}$ associated with the RAP set oscillates between the maximum and minimum values
while the occupation values corresponding to the rest of the pseudo-spins texture are only slightly affected.

This shows that a natural probe of the Rabi-Higgs mode (or other collective Rabi-like modes) would be time-resolved angle-resolved photoemision spectroscopy (tr-ARPES) from which the momentum distribution can be obtained by energy integration.  
One expects that also tunneling spectroscopy can provide information on the Rabi-like modes although since the time response of tunneling is much slower 
probably only time integrated information will be accessible. 
Explicit computation of these quantities will be presented elsewhere.

\subsection{Dynamical gapless superconductivity}
So far we have maintained the drive amplitude at relative small values $(\alpha\leq0.1)$, where the superconducting gap shows a Rabi-Higgs mode. 
Increasing $\alpha$ there is a critical value $\alpha_{c}\simeq0.3$ above which the dynamics changes completely. For $\alpha$ values close to $\alpha_{c}$, 
the details of the gap dynamics become more complicated and will be presented elsewhere. On the other hand, for large enough $\alpha$, we find a far-from-equilibrium phase of 
gapless superconductivity.

\begin{figure}[tb]
 \includegraphics[width=0.48\textwidth]{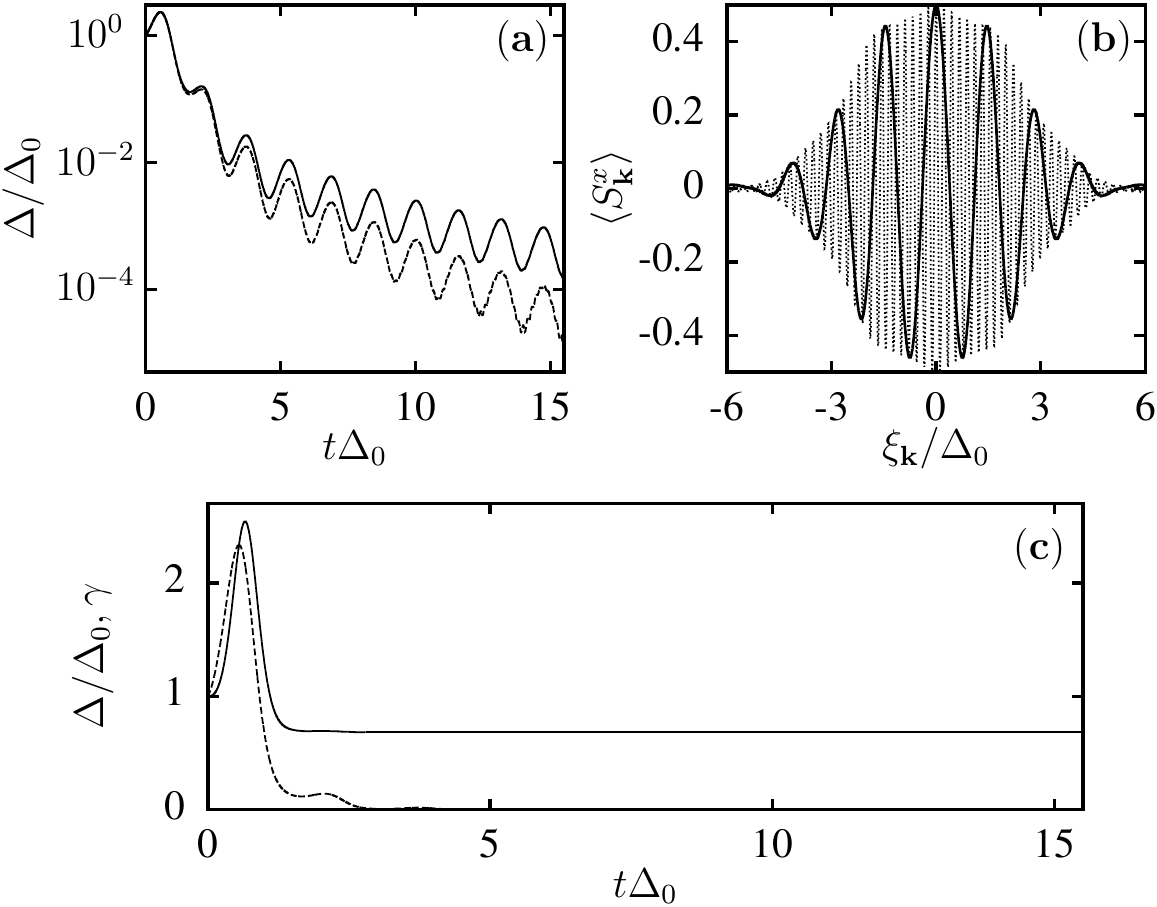}
 \caption{(a) The order parameter dynamic, in logarithmic-linear scale, for $\alpha=0.305$ and $\alpha=0.31$ in continuous and dashed line respectively. 
 (b) The $x-$component of pseudo-spins texture at $t\Delta_{0}=3$ (continuous line) and $t\Delta_{0}=15$ (dotted line) for $\alpha=0.31$ corresponding to a gapless 
 regime. (c) Superconducting gap (dashed line) and $\gamma$ (continuous line) as a function of time for $\alpha=0.31$. The $\gamma$ definition can be seen in the main text.} 
 \label{fig:gapless}
 \end{figure}

Fig.~\ref{fig:gapless}(a) shows, in logarithmic-linear scale, the gap dynamics for different values of $\alpha$. The order parameter first increases and then goes 
to zero exponentially in time showing oscillations with the drive frequency. As is shown in Fig.~\ref{fig:gapless}(b), inside this gapless phase, the $x-$component of pseudo-spins texture $\langle S_{\bm{k}}^{x}\rangle$
is non zero but its average is zero. It reflects the fact that there are pair-correlations between fermions, for instance $\langle c_{\bm{k}\uparrow}^{\dagger}(t)c_{-\bm{k}\downarrow}^{\dagger}(t)\rangle\neq0$ although
$\Delta(t)=0$, a consequence of a full pseudo-spins dephasing. In order to characterize this state, we define the quantity 
$\gamma=\beta\sum_{\bm{k}}\left\langle S_{\bm{k}}^{x}\right\rangle ^{2}$ where we choose $\beta$ such that $\gamma=1$
at $t=0$ (equilibrium). While the superconducting gap (average of $\langle S_{\bm{k}}^{x}\rangle$ essentially) goes to zero for long times, 
$\gamma$ remain as a finite constant value as is shown in Fig.~\ref{fig:gapless}(c).

 We should emphasize that, in contrast to previous works \cite{Barankov2006a,Yuzbashyan2006a,Foster2017}, this gapless regime is obtained under the presence of drive. In this 
 sense we obtain a {\em dynamical} phase of gapless superconductivity by increasing the drive strength.

\section{Application of Rabi-Higgs oscillations to the generation of dynamical phase transitions}
The spontaneous (drive-free) dynamical phases of a fermionic condensate were reviewed in the introduction. In the initial theoretical proposal \cite{Barankov2004,Barankov2006a,Yuzbashyan2006} these phases were reached after an 
interaction quench. Despite intense theoretical studies focused in the peculiarities of ultra-cold atoms \cite{Hannibal2018,Hannibal2018a} such phases have not been found experimentally so far. The situation is even worst in  solid state systems where it is difficult to imagine how large interaction quenches can be achieved. In this section we describe an alternative route to the generation of spontaneous dynamical phases based on the use of the Rabi-Higgs mode that can be applied  to both ultra-cold atoms and solid state systems.  

The idea is to apply the drive for an amount of time $t^{\star}<\tau$ i.e. smaller than a full Rabi cycle, in such a way that only a fraction of the population 
inversion of RAP has been achieved. Then the system is allowed to evolve 
spontaneously without the drive.   
In the case of phonon assisted drives generated with ISRS the lattice motion can be stopped applying 
a second pump pulse delayed from the first one by a time equal to a half integer times the drive periodicity, i.e. 
$\left(n+\frac{1}{2}\right)\mathcal{T}$, as can be easily seen solving the equations of motion for a driven harmonic oscillator. On the other hand, THz drives and drives in ultra-cold atoms can be turned on and off at will with standard experimental setups.
\subsection{Numerical Simulation}
We test this idea numerically with $\lambda$-drives switching on the drive at 
$t=0$ as before, and switching it off at $t=t^*$  during the first period of the Rabi cycle following the subsequent spontaneous out-of-equilibrium dynamics.    
Remarkably, for moderate values of $\alpha$ we obtain basically two dynamical phases, 
corresponding to relaxation or persistent oscillations of the order parameter, depending on the fraction of the Rabi cycle that has been completed. This is parameterized by $t^*/\tau$.  

\begin{figure}[tb]
\includegraphics[width=0.45\textwidth]{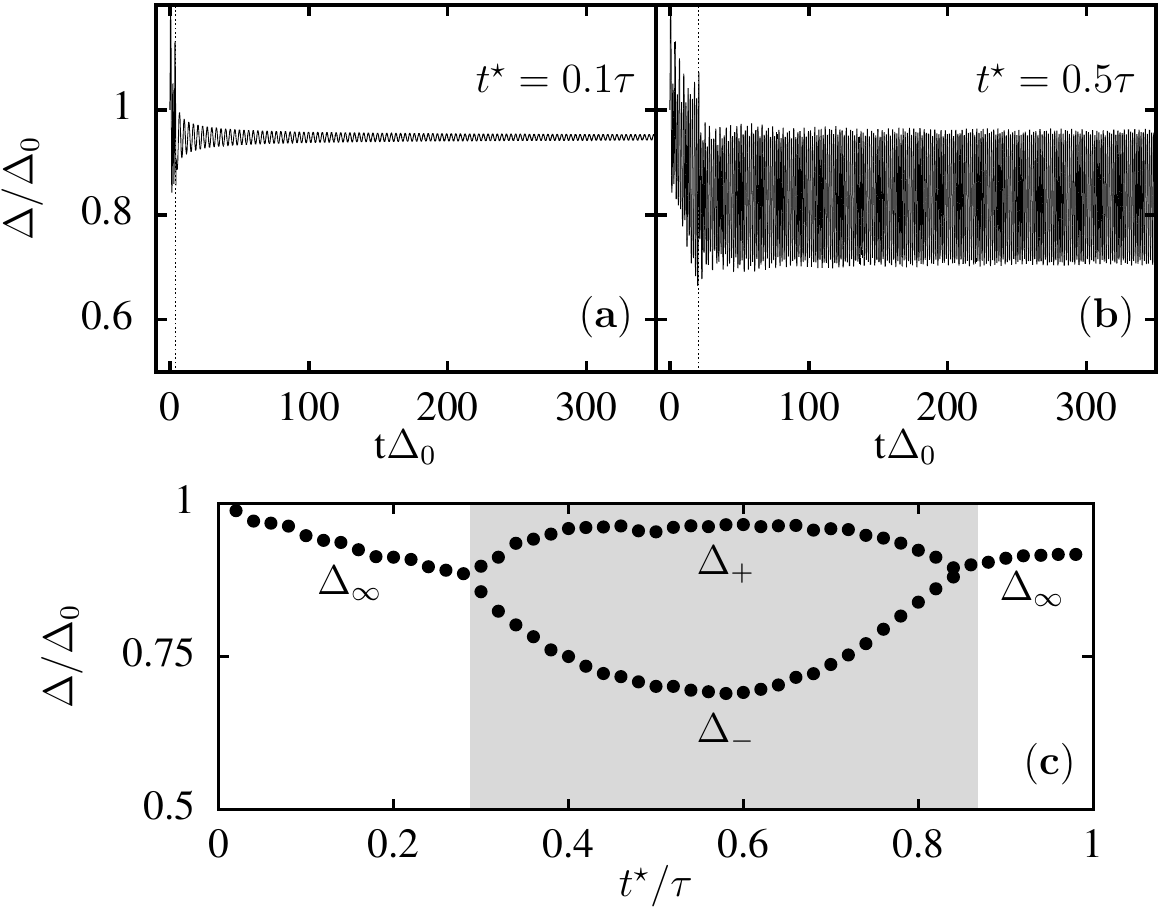}
\caption{(Top panels) Dynamics of the superconducting gap with the proposed protocol for $\alpha=0.06$ and different switch-off times $t^{\star}$. 
The corresponding values of $t^{\star}$ are indicated by a vertical dashed line and pointed in each panel as a function of the period associated with RAP $\tau\Delta_0\simeq42$. (Bottom panel) Dynamical phase diagram as a function 
of $t^{\star}/\tau$. The white regions correspond to decaying oscillations while the gray regions show persistent oscillations. We also show the gap parameters characterizing the dynamics. Phase boundaries are located at  
$t^{\star}\simeq0.3\tau$ and $t^{\star}\simeq0.87\tau$ and coincide approximately with RAP oriented perpendicular to the equilibrium field
 $\bm{b}_{\bm{k}_r}^0$.} \label{fig:dyn}
\end{figure}

 Figure~\ref{fig:dyn} shows the gap evolution for two values of $t^{\star}/\tau$. For small $t^{\star}/\tau$ the gap asymptotically approach  a constant value $\Delta_{\infty}<\Delta_{0}$ exhibiting the well known  oscillatory behavior with $1/\sqrt{t}$ decay and  frequency $2\Delta_{\infty}$ \cite{Barankov2004,Barankov2006a,Yuzbashyan2006}.
For intermediate values of $t^{\star}/\tau\sim0.5$ the gaps shows spontaneous persistent oscillations  between the two extrema $\Delta_{\pm}$ as reported before for interaction quenches \cite{Barankov2004,Barankov2006a,Yuzbashyan2006}.
The dynamical phase diagram for the proposed protocol with the long-time gap values of the gap and $\Delta_{\pm}$ values is summarized in 
the bottom panel of Fig.~\ref{fig:dyn}. In the gray region spontaneous persistent oscillations are observed while in the white region one sees the decaying
oscillations. 

In the previous section, for large values of $\alpha$, we have obtained a gapless phase under the presence of the drive (see Fig.~\ref{fig:gapless}(a)). 
It is natural to ask what is the fate of this state if the drive is turned off
at a time $t^*$ in the gapless regime.
 We checked that if $\Delta(t^*)$ is small enough the gapless regime remain.
A similar result has been obtained before in a study of the response of a
superconductor to a short but intense pump pulse in which $\Delta(t)$ is strongly suppressed before the drive-free evolution \cite{Foster2017}.
\subsection{Lax reduction analysis}
\begin{figure}[tb]
\includegraphics[width=0.5\textwidth]{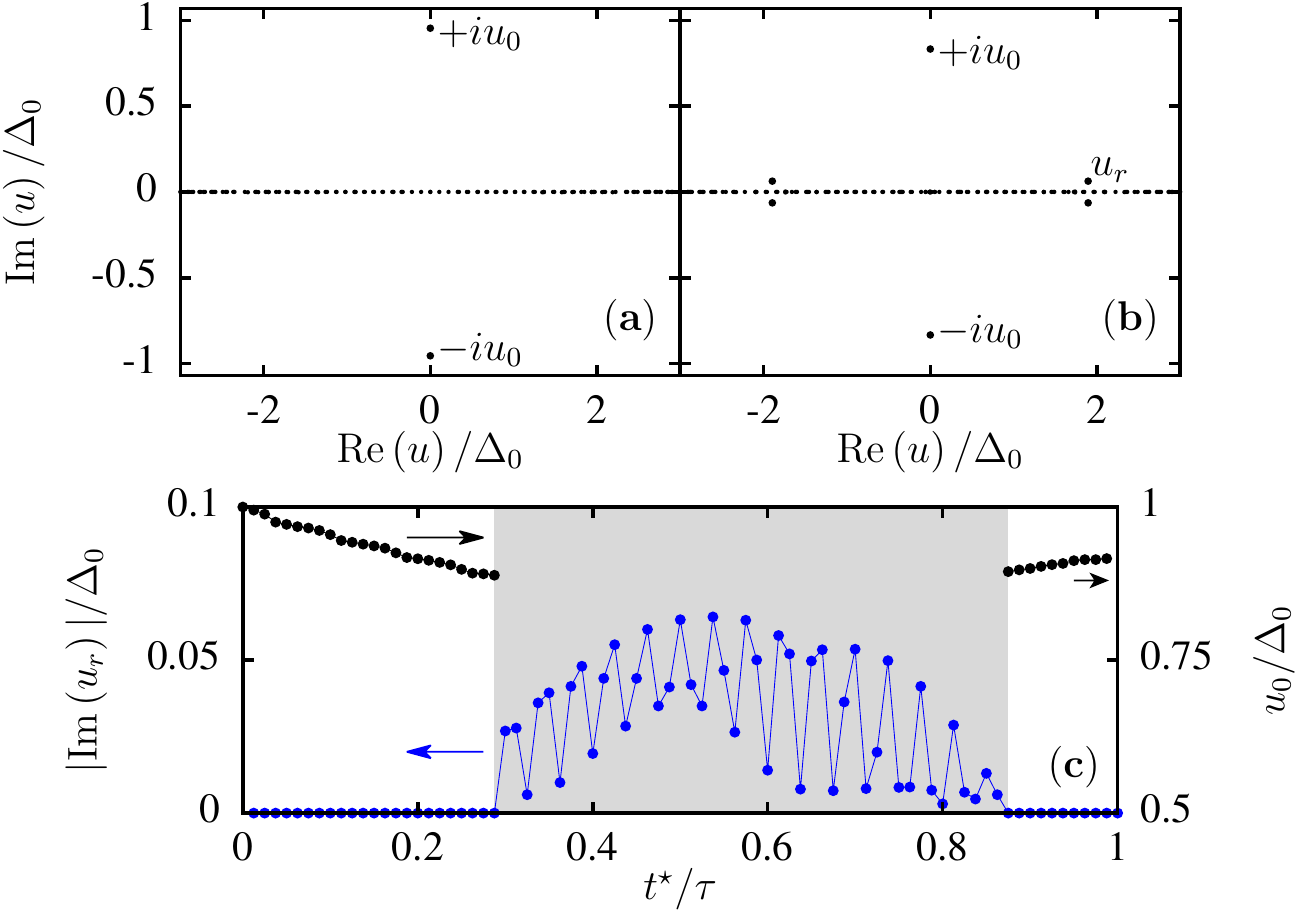}
\caption{(Color online) (Top panel) The Lax roots obtained by using the pseudo-spins texture at $t^{\star}=0.1\tau$ a) and $t^{\star}=0.5\tau$ b) with $\alpha=0.06.$ 
(Bottom panel) Imaginary part of Lax roots as a function of $t^{\star}/\tau.$ } \label{fig:l2}
\end{figure}

We verified the dynamical phase diagram of Fig.~\ref{fig:dyn} by using the Lax reduction method \cite{Yuzbashyan2005,Yuzbashyan2006}. 
The integrability of the model in the absence of drive implies that the frequency spectrum that determines the spontaneous evolution of the order parameter is determined by a set of integrals of motion.
The latter can be  evaluate at any time with the so-called Lax vector \cite{Barankov2006a,Yuzbashyan2006} 
 defined as a function of an auxiliary parameter $u$,
\begin{equation}
 \label{eq:l}
\bm{L}\left(u\right)=\bm{z}+\lambda\sum_{\bm{k}}\frac{\bm{S_{\bm{k}}}}{u-\xi_{\bm{k}}}\,.
\end{equation}
The square of the Lax vector is a conserved quantity under time evolution with the unperturbed BCS Hamiltonian and therefore its roots (in the following Lax roots) are also conserved. Since the square of the Lax vector is non-negative, all roots should come in complex-conjugated pairs. Also, 
since $S^{x}_{\bm{k}}=S^{x}_{-\bm{k}}$ and $S^{z}_{\bm{k}}=-S^{z}_{-\bm{k}}$ holds at all times, where $\bm{k}$ and $-\bm{k}$ label the quasi-particle states with 
energy $\xi_{\bm{k}}$ and $\xi_{\bm{-k}}=-\xi_{\bm{k}}$, respectively, it is easy to see  that if $u$ is a Lax root $-u$ is also a root (see top panels of Fig.~(\ref{fig:l2})).
In Ref. [\onlinecite{Yuzbashyan2006}] it was shown that the dynamics of the order parameter is related to the number $m$ of isolated pairs of complex-conjugated Lax roots. Indeed, $\Delta\left(t\right)$ shows persistent oscillations at long times with $k$ different 
frequencies if $m>1$  while $\Delta\left(t\right)\rightarrow\Delta_{\infty}$ (damped oscillations) if $m=1$. Here, $k$ is equal to the integer part of $m/2$. 

To check the numerical results we constructed the Lax vector and compute its roots from the instantaneous pseudo-spin texture at time $t^{\star}$, i.e. with the initial condition of the subsequent free evolution. As an example,  we show  
in the top panels of Fig.~\ref{fig:l2} the Lax roots in the complex plane for the same values of $t^{\star}/\tau$ that were used in top panels of Fig.~\ref{fig:dyn}. Consistently with the numerical results, 
for $t^{\star}=0.1\tau$ we obtain one pair of isolated Lax roots ($m=1$, damped dynamics) and for $t^{\star}=0.5\tau$ we obtain $m=3$ (persistent oscillations).

For any value of $t^{\star}$ we have one pair of purely imaginary isolated Lax roots $\pm iu_{0}$ plus a continuum of doubly degene\-rated roots on the real axis.
In the damped regime, the absolute value of the purely imaginary roots determine the 
asymptotic value of the superconducting gap, $\Delta_{\infty}=u_{0}$ \cite{Barankov2006a}. In the persistent oscillatory phase two extra pairs of isolated roots $\pm u_{r} \left(\pm u_{r}^{*}\right)$ appear (see Fig.~\ref{fig:l2}(b)). 
We find numerically that $\mathrm{Re} \left(u_{r}\right)$ is close to $\xi_{\bm{k}_{r}}$ and the imaginary part is finite. 
Varying $t^{\star}/\tau$, the  change of behavior to the damped regime 
is determined by $\mathrm{Im}\left(u_{r}\right)\rightarrow 0$. 
In the bottom panel of 
Fig.~\ref{fig:l2} we plot the imaginary part of these Lax roots $u_{r}$ as a function of $t^{\star}$ which reproduces the phase diagram obtained numerically [c.f. Fig.~(\ref{fig:dyn})]. We also show $u_0$ vs.  
$t^{\star}$ which accounts for  the values of $\Delta_{\infty}$ shown in Fig.~\ref{fig:dyn}.

\section{Conclusions}
\label{sec:conc}
We have shown that a superconductor subject to a perio\-dic harmonic drive in the collisionless regime perform Rabi-Higgs oscillations with a frequency that can be controlled by the strength of the drive. 
We argued that there are several solid state and ultra-cold atom routes to realize the drive. 

Periodic drives have been considered before\cite{Tsuji2015,Sentef2017,Cea2018}. Our results go beyond previous computations of third-order susceptibilities \cite{Xi2013,Tsuji2015,Cea2018} by considering the full dynamical non-linear response 
beyond perturbation theory. Ref.~\cite{Sentef2017} finds a slow oscillation of the superconducting order parameter under a drive, in a model in which superconductivity is close to a charge density wave instability.  Then the system oscillates periodically between different orders with a frequency much slower than the drive. Our results apply instead to the more general case of a BCS superconductor which is not close to another order and the oscillations are between an equilibrium superconductor and a highly excited superconductor.

In real solid-state and cold-atom systems, terms in the Hamiltonian beyond BCS mean-field will produce collisions and tend to relax the out of equilibrium populations. It is hard to predict in advance which system can fulfill the requirement that the coherence time is long enough to see the 
Rabi-Higgs oscillations. In general low pump fluence will reduce heating but produce slower oscillations which may require longer coherence time to be seen. Therefore, a tread off  should be found which will depend on the
specific pump mechanism and material. 
In the present mechanism the drive frequency has to be larger than $2 \Delta_0$ but be of the same order.
Keeping $\alpha$ and the ratio $\omega_d/ \Delta_0$ fixed, the Rabi frequency scales linearly
with the gap [cf. Eq.~(\ref{eq:delta1}) and Eq.~(\ref{eq:w0})]:
\begin{equation}
\label{eq:wr}
\omega_R= \alpha\left|1+\chi_{\Delta,\Delta}(\omega)\right|\Delta_{0} \sqrt{1-\left(\frac{2 \Delta_0
}{\omega_d}\right)^2},
\end{equation}
thus lowering the gap makes the Rabi oscillation longer $\tau \sim 1/\Delta_{0}$. On the
other hand, coherence times usually scale faster with quasiparticle energy. For
example in a Fermi liquid, decay due to electron-electron interaction scales as $1/\omega^{2}$.  Therefore, working with materials with small gaps can be beneficial as coherence times can be made longer than the Rabi time one would like to measure. Taking into account that the system is not a Fermi liquid but a superconductor the situation is, of course, even better. 
In this regards it is encouraging to notice that Rabi oscillations are not unprecedented in the solid state but have a long history in semiconductors\cite{Cundiff1994,Furst1997,Giessen1999} and are part of the 
modern toolbox of solid-state quantum technologies\cite{Cole2001,Stievater2001,Htoon2002,Zrenner2002}. 
 
We have discussed drives which preserve the symmetry of the
lattice and  couple to the Higgs modes. It is also possible to apply
drives that do not preserve the symmetry of the lattice and couple,
for example, to $B_{1g}$ and
 $B_{2g}$ charge fluctuations in the case of a square lattice. These
drives will lead to Rabi-charge oscillations entirely analogous to the
Rabi-Higgs oscillations. An advantage of these  modes is that being
Raman like they modulate the optical properties of the material as the
analogous spontaneous modes do \cite{Mansart2013,Lorenzana2013}.
Therefore, they can be studied by optical means for example by pumping
in the THz range and probing through differential reflectivity with
visible light.

As an application we have shown how the induced far from equilibrium state
can be used to observe the so far elusive spontaneous dynamical phases
predicted by theory \cite{Barankov2004,Barankov2006a,Yuzbashyan2006}.
We showed that Lax reduction method provides a general quantitative
tool to discriminate the different dynamical phases.
The original proposals were based on a very strong perturbation
applied in a short time. Instead, our proposal is based on a gentle
perturbation applied for a moderate time exploiting the properties of
collisionless or nearly collisionless condensates. This opens a window
of opportunity to make these phases accessible even in a solid state
setting.

Our work is an example of quantum control of a condensate wave function.
The population of a set of quasi-particles is periodically inverted.
The strong analogy with NMR and solid state quantum-control 
\cite{Cundiff1994,Furst1997,Giessen1999,Cole2001,Stievater2001,Htoon2002,Zrenner2002} pave the way to
replicate well known NMR and optical protocols in fermionic superfluids and
observe other fascinating phenomena as for example Hahn
echoes \cite{Hahn1950} or soliton propagation\cite{Giessen1998}.

\begin{acknowledgments}
J.L. is very much in debt with   Nicolas Bergeal, J\'er\^ome Lesueur, Gabriel Lemari\'e, Claudio Castellani and Lara Benfatto for useful conversations.     We acknowledge financial support 
from Italian MAECI and Argentinian MINCYT through  bilateral project AR17MO7 
and  Italian MAECI thought collaborative project SUPERTOP-PGR04879. 
We acknowledge financial support from ANPCyT (grants PICTs 2013-1045 and 2016-0791),  CONICET (grant PIP 11220150100506) and SeCyT-UNCuyo C (grant 06/C526).
\end{acknowledgments}



\end{document}